\RequirePackage{etex}
\documentclass[superscriptaddress,12pt]{amsart}
\usepackage{amsmath, amsthm, amssymb,  mathrsfs, epsfig}
\usepackage{dcpic, pictex, pinlabel}
\usepackage{hyperref}
\usepackage{color}

\newcommand{\A}{\alpha}
\newcommand{\B}{\beta}

\theoremstyle{definition}
\newtheorem{axiom}{Axiom}
\newtheorem{thm}{Theorem}

\newtheorem{lem}{Lemma}

\newtheorem{cor}{Corollary}

\newtheorem{rem}{Remark}
\newtheorem{definition}{Definition}
\newtheorem*{sketch}{Sketch of Proof}

\numberwithin{equation}{section} \makeatletter
\renewenvironment{proof}[1][\proofname]{\par
    \pushQED{\qed}%
    \normalfont \topsep6\p@\@plus6\p@ \labelsep1em\relax
    \trivlist
    \item[\hskip\labelsep\indent
        \bfseries #1]\ignorespaces
}{%
    \popQED\endtrivlist\@endpefalse
} \makeatother
\renewcommand{\proofname}{Proof}

\begin{document}

\title{Complexity Classes as Mathematical Axioms \textrm{II}}

\author{Shawn X. Cui$^{1}$, Michael H. Freedman$^{1,2}$, and Zhenghan Wang$^{1,2}$}

\address{$^1$Department of Mathematics\\University of California\\Santa Barbara, CA 93106}
\email{xingshan@math.ucsb.edu, zhenghwa@math.ucsb.edu}
\address{$^2$Microsoft Research, Station Q\\ University of California\\ Santa Barbara, CA 93106}
\email{michaelf@microsoft.com, zhenghwa@microsoft.com}

\thanks{S.-X. C and Z.W. are partially supported by NSF DMS 1108736. We thank two anonymous referees for their suggestions on the definition of $NDQC1$.}


\date{}

\begin{abstract}
The second author previously discussed how classical complexity separation conjectures, we call them \lq\lq axioms", have implications in three manifold topology: polynomial length strings of operations which preserve certain Jones polynomial evaluations cannot produce exponential simplifications of link diagrams. In this paper, we continue this theme, exploring now more subtle separation axioms for quantum complexity classes. Surprisingly, we now find that similar strings of operations are unable to effect  even linear simplifications of the diagrams.
\end{abstract}
\maketitle

\section{Introduction}

Evaluation of the Jones polynomial at $\omega_r = e^{\frac{2\pi i}{r}}$ is known to be $\sharp P$ hard for $r\geq 5$ except $r=6$ \cite{Vertigan,Vertigan2}. Following \cite{Freedman}, we consider strings of Dehn surgeries designed to be easily (polynomially) describable and not to alter the  $\omega_r$-Jones evaluation. In the earlier paper \cite{Freedman}, it was shown that polynomial length strings of such Dehn surgeries cannot effect an exponential simplification of the general link diagrams without contradicting standard conjectures---they were called \lq\lq axioms"---regarding the separation of classical complexity classes. In particular, it was shown that $``P^{PP}$ not in $NP"$ would be contradicted by exponential simplification of diagrams\footnote{The new diagrams have crossing numbers or girths that are logarithmic in the crossing numbers or girths of the original diagrams via a sequence of polynomially many moves.}. In this paper, we continue this theme, exploring now more subtle separation axioms for quantum complexity classes. Moreover, we generalize the Dehn surgeries in \cite{Freedman} so that we have more flexibility to change the link while preserving its Jones evaluations. Surprisingly, we now find that even the generalized strings are unable to effect  even linear simplifications of the diagrams without contradicting the separation \lq\lq axiom" that $BQP$ is not in $NDQC1$---a complexity class that we will define in Section \ref{axiom}.  For more on the history of the question and background on complexity theory, see \cite{Freedman}.

$BQP$ is the class of problems that the quantum circuit model can solve efficiently.  One complete problem for $BQP$ is the approximation of the $\omega_r$-Jones evaluation at $r=5$ \cite{Freedman03}.  Detailed analysis of the approximation revealed two subtleties:  the efficient approximation is an additive approximation and the algorithm depends on the presentation of links as plat closures \cite{BFLW,SJ}.  Additive approximation turns out to be very sensitive to the presentation of links.  As shown in \cite{SJ}, the approximation of the $\omega_r$-Jones evaluation at $r=5$ for trace closure is complete for DQC1---a complexity class conjecturally strictly weaker than $BQP$.  Our sharpening of the result in \cite{Freedman} is an elucidation of the dependence of the approximation on the presentation of links.

Throughout this paper, we use the following notation:
$r$ is an integer $\geq 5$ except $r=6$, and $d = 2\cos(\frac{\pi}{r})$. For a braid $\sigma \in B_{2n}$, $\hat{\sigma}^{plat}$ is the plat closure. For $\sigma \in B_n,$ $\hat{\sigma}^{tr} $ is the trace closure. $J(\hat{\sigma}^{plat};r)$ and $J(\hat{\sigma}^{tr};r)$ are the Jones polynomial at the $r$-th root of unity $\omega_r$ of the plat closure and trace closure of $\sigma$, respectively.

The rest of the paper is organized as follows. In Section \ref{r-equ}, we define an equivalence relation on links and a distance between two equivalent link diagrams. Section \ref{app jones} consists of background materials on approximating the Jones polynomial at the root of unity $\omega_r$. In Section \ref{axiom}, we introduce three \lq\lq axioms" regarding the separation of complexity classes. Finally in Section \ref{main thm}, we obtain three theorems, each of which follows from one of the axioms in Section \ref{axiom}.

\section{Surgeries and Axioms}

\subsection{An equivalence relation and a distance} \label{r-equ}

First we introduce an equivalence relation on links which generalizes the relation $\sim_r$ in \cite{Freedman}. Links and link diagrams in this paper are oriented and framed.

For a link diagram $D$ in the $x$-$z$ plane with the blackboard framing, we define its {\it{girth}} $g(D)$ to be the maximum over all $z_0$ of the cardinality of the set $D \cap \{z = z_0\}$, and define its {\it complexity} $c(D)$ to be the number of crossings of $D$. Intuitively, $g(D)$ measures how wide $D$ spreads along the $x$-axis direction, while $c(D)$ measures the \lq\lq area" of the diagram. Note that the definition of $c(D)$ here is different from that in \cite{Freedman} since we will not deal with the maxima and minima of a diagram. For a link $L$, we define its girth $g(L)$ = min$\{g(D)\ |$ $D$ is a diagram of $L$ $\}$. The complexity $c(L)$ is defined similarly, i.e. $c(L)$ = min$\{c(D)\ |$ $D$ is a diagram of $L$ $\}$.

Given a link $L$ in $S^3$, recall in \cite{Freedman} that the $\frac{\pm1}{4r}$-Dehn surgery on $L$ is defined as follows. Consider $L \amalg U$, where $U$ is an unknot disjoint from $L$, and then perform a $\frac{\pm1}{4r}$-Dehn surgery on $U$ to change $L$ into a new link $L'$. It turns out that $L'$ has the same $\omega_r$-Jones evaluation as $L$. For example, if $U$ bounds a standard disk and this disk meets $L$ transversally in $n$ points, then the surgery will introduce $4rn(n-1)$ crossings to $L$.

Let $\Gamma(4r)$ be the principal congruence subgroup of level $4r$ of $\textrm{SL}(2,\mathbb{Z})$, i.e. the kernel of the group homomorphism $\textrm{SL}(2,\mathbb{Z}) \longrightarrow \textrm{SL}(2, \mathbb{Z}/4r\mathbb{Z})$ sending a matrix $A$ to $A\;\; (\textrm{mod}\;4r)$ entry-wise.

Now we generalize the $\frac{\pm1}{4r}$-Dehn surgery process. Consider a pair $(M,L)$, where $M$ is an oriented closed $3$-manifold and $L$ is an oriented framed link in $M$. For any knot $K$ disjoint from $L$ in $M$, performing a Dehn surgery of $M$ along $K$ via a map $\phi \in \Gamma(4r)$ results in a new closed $3$-manifold denoted by $M_{K,\phi}$. Passing from $(M, L)$ to  $(M_{K,\phi}, L)$ is what we call a {\it $\Gamma_{r}$-Dehn surgery} on $(M,L)$; it generates an equivalence relation on pairs $(M,L)$ similar to the \lq\lq congruence" studied by Lackenby \cite{Lackenby} and Gilmer \cite{Gilmer}.

Given a link $L$ in $S^3$, a sequence of $\Gamma_{r}$-Dehn surgeries leads to a pair  $(M,L)$ of a $3$-manifold $M$ with a link $L$ inside, where $M$ may not be $S^3$. Since we are only interested here in links in $S^3$, we want the manifold $M$ to be diffeomorphic to $S^3.$  If this is the case, we need to construct a diffeomorphism from $M$ to $S^3$ which transforms $L$ to a new link $L'$ in $S^3$.  However, it is not presently known if there is an {\it{efficient}} (polynomial time) algorithm for recognizing the $3$-sphere.  But by \cite{SCHLEIMER}, verifying whether a given closed $3$-manifold is $S^3$ is a problem in $NP$. So there is a polynomial size certificate to verify if a manifold is diffeomorphic to $S^3.$ Moreover, by \cite{SCHLEIMER}, one may actually construct such a diffeomorphism if the certificate gives \lq\lq yes" output. Let us refer to this process as {\it $S^3$ recognition}.

\begin{definition}
Two oriented framed links $L$ and $L'$ are called $r$-equivalent, which is denoted by $L \sim_r L',$ if there is a sequence of operations consisting of $\Gamma_{r}$-Dehn surgeries and $S^3$ recognitions to pass from the pair $(S^3,L)$ to $(S^3,L').$   Two oriented framed link diagrams $D$ and $D'$ are $r$-equivalent if the links that they represent are $r$-equivalent.
\end{definition}

Next we want to define a distance between two $r$-equivalent diagrams $D$ and $D'$. Roughly speaking, the distance is the minimal length of the sequence of operations consisting of Reidemeister moves, $\Gamma_{r}$-Dehn surgeries, and $S^3$ recognitions. However, the knots $K_1,\cdots,K_n$ and the regluing matrices $F_1, \cdots, F_n$ in $\Gamma(4r)$ along which we do the surgeries should be efficiently describable, i.e. they should not be too complicated. Moreover, the complexity of $S^3$ recognition process would depend on the complexity of the Dehn surgery description. Considering these factors, we will assign a {\it{weighted}} distance to the transition from $D$, a diagram for $(S^3,L)$, to $D'$ for $(S^3, L').$ To do so, we use a larger, intermediate, diagram, $\widetilde{D}'(S^3, L'),$ a literal $\Gamma_{r}$-Dehn surgery diagram for $(S^3, L'),$ consisting of $L$ and the components $K_1,\cdots,K_n$ labeled by $F_1, \cdots, F_n$, which designate the $\Gamma_{r}$-Dehn surgeries.

\begin{definition}
For two $r$-equivalent diagrams $D$ and $D'$, the $r$-distance $dist_r(D,D')$ is the minimum of $c+b+\gamma$, where $c$ is the total number of crossings (of all types) in $\widetilde{D}'(S^3, L'),$ $b$ is the number of bits needed to write the integral entries of matrices $\{ F_1, \cdots ,  F_n\},$ and $\gamma$ is the number of Reidemeister moves required to take the image of $L$ in the resulting manifold, after sphere recognition has been applied, and transform it to $L'$. The minimum is taken over all possible $\Gamma_{r}$-Dehn surgery diagrams and all subsequent sequences of Reidemeister moves.
\end{definition}

\begin{rem}
The explicit form of $dist_r$ is not relevant. What is important is that the number of computational steps to pass from $D$ to $D'$ is not larger than a polynomial in  $dist_r(D,D')$. The number of computational steps of each $\Gamma_{r}$-Dehn surgery and each Reidemeister move is clearly bounded by a polynomial in the weights they contribute to $dist_r(D,D')$. After the $\Gamma_{r}$-Dehn surgeries, one can construct a triangulation of the resulting manifold. And the number of simplices in the triangulation is less than $poly(b+c).$ So the number of computational steps of $S^3$ recognition is also a polynomial in $b+c$, hence a polynomial in $dist_r(D,D')$.

It is straightforward to pass in polynomial time between any of the common methods for describing a $3$-manifold. For the case here, one just imbeds the link diagram in the $1$-skeleton of some triangulation $\Delta_1$ of $S^3$ and then extends $\Delta_0$---the restriction of $\Delta_1$ to the link complement---to a triangulation $\Delta$ of the Dehn surgered manifold, $\Delta=\Delta_0\cup \Delta_{\textrm{solid tori}}$.  To match the triangulation $\Delta_0$ on the link complement with the triangulation on the solid tori $\Delta_{\textrm{solid tori}}$, both may have to be refined.  The amount of refinement is polynomial in the entries of the gluing matrices.
\end{rem}

The following lemma says that the Dehn surgery preserves the absolute value of the Jones polynomial at the $r$-th root of unity $\omega_r$.

\begin{lem}\cite{Freedman}
If $D \sim_r D', $ then $|J(D;r)| = |J(D';r)|.$

\begin{proof}
This lemma was from \cite{Freedman}. Here we prove it in the more general case. The connection between $\omega_r$-Jones evaluation and the Dehn surgery is the $SU(2)$-Reshetikhin-Turaev topological quantum field theory (TQFT) at level $k=r-2$, which we call $SU(2)_{k}$-TQFT and denote by $(V_k, Z_k)$ . Let $L$ be the link represented by $D$. By \cite{TR}, $J(L,r) = Z_k(S^3, L),$ the partition function of $(S^3,L)$ in the $SU(2)_{k}$-TQFT.

Now we prove that the Dehn surgery we defined preserves the partition function of $(M,L)$ up to a phase, where $M$ is an oriented closed $3$-manifold. Let $K$ be the knot along which we do the surgery and let $T$ be the torus which bounds a solid torus neighborhood $N$ of $K$. Then $V_k(T) \cong \mathbb{C}^{k+1}, $ where $V_k$ is the $\textrm{SU}(2)_k$-modular functor. Let $\phi \in \Gamma(4r)$ be the gluing map. Then we have the formula $Z_k(M_{K,\phi},L) = \langle Z_k(\overline{M\backslash N},L) | V_k(\phi)(Z_k(N))\rangle. $ By \cite{NS}, $\Gamma(4r)$ is contained in the kernel of the (projective) modular representation corresponding to $SU(2)_{k}$-TQFT. Thus $V_k(\phi)$ acts as identity up to a phase. So $|Z_k(M_{K,\phi},L)| = |\langle Z_k(M-N,L) | Z_k(N)\rangle| = |Z_k(M,L)|.$

The diffeomorphism from a manifold $M$ to $S^3$ preserves the partition function up to a phase. So we have $|Z_k(S^3,L)| = |Z_k(S^3, L')|,$ which implies $|J(L,r)| = |J(L',r)|.$
\end{proof}
\end{lem}

\subsection{Approximating the Jones polynomial}\label{app jones}
The following theorem can be found in various references, e.g. \cite{Freedman03,Aha09,WY,BFLW,Wang,Kuperberg09,aharonov2011,freedman2002,freedman20022}, which says approximating the Jones polynomial of the plat closure of a braid at the $r$-th root of unity $\omega_r$ is $BQP$-complete for $r\geq 5$ except $r=6$.

\begin{thm}\label{BQP}\cite{Freedman03,Aha09,WY,BFLW,Wang,Kuperberg09,aharonov2011,freedman2002,freedman20022}
There is an efficient classical algorithm which, given a braid $\sigma \in B_{2n} $ of length $m$ and an error threshold $\varepsilon > 0 $ as input, outputs a description of a quantum circuit $U_{\sigma,\varepsilon}$ of size $poly(n,m,\frac{1}{\varepsilon}).$ This quantum circuit computes a random variable $0 \leq Z(\sigma)\leq 1,$ such that
$$
Pr\bigg\{\big |\frac{|J(\hat{\sigma}^{plat};r)|}{d^{n}} - Z(\sigma)\big | < \varepsilon\bigg\} > \frac{3}{4}.
$$

Moreover, the problem of approximating $\frac{|J(\hat{\sigma}^{plat};r)|}{d^n}$ for a braid $\sigma \in B_{2n}$ is $BQP$-complete.
\end{thm}

In \cite{SJ}, the authors showed that for $r = 5$, approximating the Jones polynomial of the trace closure of a braid at the $r$-th root of unity is $DQC1$-complete. In \cite{JW}, it was further shown that actually approximating the Jones polynomial of the trace closure at any $r$-th root of unity is a $DQC1$ problem.  Here $DQC1$ is the set of problems which can be solved efficiently by a one clean qubit quantum computer \cite{KL}. One clean qubit means the initial state consists of a single qubit in the pure state $|0\rangle,$ and $n$ qubits in a maximally mixed state. This is described by the density matrix
$$
\rho = |0\rangle \langle 0 | \otimes \frac{I}{2^n}
$$

Then we can apply a unitary evolution on these $(n+1)$ qubits and measure the clean qubit in the computational basis. The probability of measuring $|0\rangle$ is

\begin{equation} \label{equation}
p_0 = 2^{-n} Tr\{(|0\rangle \langle 0 | \otimes I) U (|0\rangle \langle 0 | \otimes) U^{\dag}\}
\end{equation}

For more detailed discussion of the $DQC1$ model, see \cite{KL, SJ}.

\begin{thm} \label{DQC1} \cite{SJ, JW}

There is an efficient classical algorithm which, given a braid $\sigma \in B_{n} $ of length $m$ and an error threshold $\varepsilon' > 0 $ as input, outputs a description of a quantum circuit $U_{\sigma,\varepsilon'}$ of size $poly(n,m,\frac{1}{\varepsilon'}).$ This quantum circuit computes a random variable $0 \leq Z(\sigma)\leq 1$ in the one clean qubit model such that
$$
Pr\bigg\{\big |\frac{|J(\hat{\sigma}^{tr};r)|}{d^{n}} - Z(\sigma)\big | < \varepsilon'\bigg\} > \frac{3}{4}.
$$
\end{thm}

\begin{thm}\cite{SJ}
When $r=5$, the problem of approximating $\frac{|J(\hat{\sigma}^{tr};r)|}{d^n}$  for a braid $\sigma \in B_n$  is $DQC1$-complete.

\begin{sketch}
The key point is that approximating the normalized trace of a circuit is $DQC1$-complete \cite{Shepherd}. We know that the Jones polynomial of a braid closure is equal to a weighted trace of the braid under the Jones representation. Thus modulo some technical details such as how to encode the Jones representation into a circuit, the theorem is plausible. We now sketch the proof that approximating the normalized trace, $\frac{tr(U)}{2^n}$, of a circuit $U$ on $n$ qubits is a $DQC1$-complete problem.

We first show approximating the normalized trace is in $DQC1$. For a circuit $U$ of $n$ qubits and a pure state $|\psi\rangle$, there is a standard way to approximate $\langle \psi | U |\psi\rangle$, which is called the {\it Hadamard test} and is shown below.

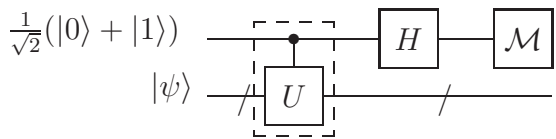
\setlength{\unitlength}{0.030in}
\begin{picture}(160,40)(-39,0)

\put(20,10){\line(1,0){10}}
\put(40,10){\line(1,0){40}}
\put(20,20){\line(1,0){30}}
\put(60,20){\line(1,0){10}}

\put(35,20){\line(0,-1){5}}

\put(35,20){\circle*{2}}

\put(30,5){\framebox(10,10){$U$}}
\put(50,15){\framebox(10,10){$H$}}
\put(70,15){\framebox(10,10){$\mathcal{M}$}}
\put(28,3){\dashbox{2}(14,20)}

\put(-15,20){$\frac{1}{\sqrt{2}}(|0\rangle + |1\rangle)$}
\put(10,10){$|\psi\rangle$}
\put(25,8){$/$}
\put(60,8){$/$}

\end{picture}

In the circuit above, a horizontal line with a slash through it represents multiple qubits. $U$ is the $n$-qubit gate and $H$ is the Hadamard gate. $\mathcal{M}$ is the measurement of the first qubit in the computational basis. Moveover, the gate in the dashed box is called Controlled-$U$ gate denoted by $\bigwedge(U).$ It's an $(n+1)$-qubit gate defined by the maps:
$$\bigwedge(U): |i\rangle|\psi\rangle \longrightarrow  |i\rangle U^i|\psi\rangle.$$

In other words, the first qubit acts as a control bit. The gate $U$ will be applied to the other $n$ qubits only if the first qubit is $1$.

The following short computation shows that the probability of obtaining $|0\rangle$ is
\begin{align*}
\widetilde{p_0} & = \big |(|0\rangle\langle 0 | \otimes Id)(H \otimes Id)\bigwedge(U)(\frac{|0\rangle+|1\rangle}{\sqrt{2}} \otimes |\psi\rangle)\big |^2 \\
                & = \big | \frac{1+U}{2} |\psi\rangle \big |^2 \\
                & =\frac{1 + Re (\langle \psi | U |\psi\rangle)}{2}.
\end{align*}

Similarly, if the control bit is initialized with $\frac{1}{\sqrt{2}}(|0\rangle - i|1\rangle),$ then the probability to obtain $|0\rangle$ is $\frac{1 + Im (\langle \psi | U |\psi\rangle)}{2}$.

Now suppose $|\psi\rangle$ is in a maximally mixed state, then running the Hadamard test to it gives the probability
$$
\widetilde{p_0} = \sum\limits_{x \in \{0,1\}^n} \frac{1}{2^n}\frac{1+ Re(\langle x | U | x \rangle)}{2} = \frac{1}{2} + \frac{Re(Tr \, U)}{2^{n+1}}.
$$

Notice that the maximally mixed state, together with a clean qubit, is exactly the input of the one clean qubit model. We can use the one clean qubit as the control qubit in the Hadamard test to convert the circuit $U$ into Controlled-$U$. Therefore, the one qubit clean model can approximate the normalized trace.

Next we show that approximating the normalized trace is $DQC1$-complete. This should be more or less clear from Equation \ref{equation}, namely after applying an $(n+1)$-qubit gate $U$ to the density matrix $|0\rangle \langle 0| \otimes \frac{I}{2^n}$ followed by the measurement of the clean qubit, the probability to obtain $|0\rangle$ is $p_0 = 2^{-n} Tr\{(|0\rangle \langle 0 | \otimes I) U (|0\rangle \langle 0 | \otimes) U^{\dag}\}$. Note that $(|0\rangle \langle 0 | \otimes I) U (|0\rangle \langle 0 | \otimes) U^{\dag}$ is not a unitary transformation so we cannot approximate the trace directly. However, the following circuit $U'$ can be easily constructed, and one can check that $p_0 = \frac{tr(U')}{2^{n+2}}.$ Also note that $U'$ is an $(n+3)$-qubit gate, thus we can approximate the normalized trace $\frac{tr(U')}{2^{n+3}}.$ This shows that approximating the normalized trace is $DQC1$-complete.

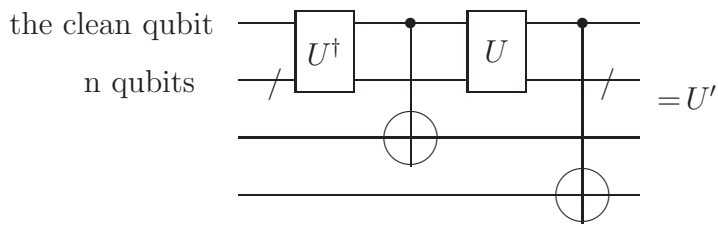
\setlength{\unitlength}{0.030in}
\begin{picture}(160,50)(-39,0)

\put(10,10){\line(1,0){70}}
\put(10,20){\line(1,0){70}}
\put(10,30){\line(1,0){10}}
\put(30,30){\line(1,0){20}}
\put(60,30){\line(1,0){20}}
\put(10,40){\line(1,0){10}}
\put(30,40){\line(1,0){20}}
\put(60,40){\line(1,0){20}}

\put(40,40){\line(0,-1){25}}
\put(70,40){\line(0,-1){35}}

\put(40,40){\circle*{2}}
\put(70,40){\circle*{2}}

\put(70,10){\circle{10}}
\put(40,20){\circle{10}}

\put(20,28){\framebox(10,14){$U^{\dag}$}}
\put(50,28){\framebox(10,14){$U$}}
\put(88,25){$U'$}
\put(83,25){=}

\put(15,28){$/$}
\put(73,28){$/$}
\put(-17,28){$\textrm{n qubits}$}
\put(-30,38){$\textrm{the clean qubit}$}
\end{picture}
\qed
\end{sketch}
\end{thm}

\begin{rem}
For $r > 6,$  we cannot find a proof of the $DQC1$-hardness of approximating $\frac{|J(\hat{\sigma}^{tr};r)|}{d^n}$ for a braid $\sigma \in B_n$ in the literature. However, its containment in $DQC1$ is sufficient for our purpose.  Therefore, our theorems below hold for all $r > 6$ and $r = 5.$
\end{rem}

\subsection{Complexity Classes as Axioms}\label{axiom}

It is easy to show that $DQC1 \subset BQP$. But it's not known whether this inclusion is strict or not. Another generally accepted conjecture is that $BQP \nsubseteq NP.$ We are going to give a stronger assumption.

We define a new complexity class $NDQC1$, which informally is the composition of $NP$ and $DQC1$.  Intuitively, a problem $X$ in $NDQC1$ can be solved by solving two related problems, the first one is in $NP$ and the second one in $DQC1$.  We model our definition of $NDQC1$ on the definition of $NP$ using a verifier: instead of a $P$ verifier, the verifier for our complexity class is a $DQC1$ machine.  Note that $DQC1$ as defined in \cite{KL} contains $P$.

\begin{definition}
The complexity class $NDQC1$ is defined as follows:  an instance $x$ of an $NDQC1$ problem $X$ has a \lq\lq Yes" answer if there exists a polynomial-length certificate $y(x)$ such that the verifier, which is a $DQC1$ algorithm, accepts the input $(x,y(x))$ with probability $>\frac{2}{3}$; and the instance has a \lq\lq No" answer if, for all $y(x)$ of the specified length, the verifier rejects the input $(x,y(x))$ with probability $>\frac{2}{3}$.
\end{definition}

Obviously, $NDQC1$ contains both $NP$ and $DQC1$. But whether or not $NDQC1$ contains $BQP$ is open.

\begin{axiom} \label{A1}
$BQP \nsubseteq NDQC1$
\end{axiom}

\begin{axiom} \label{A2}
$BQP \nsubseteq NP$
\end{axiom}

\begin{axiom} \label{A3}
$BQP \nsubseteq DQC1$
\end{axiom}

\begin{rem}
Note that Axiom \ref{A1} is potentially stronger than Axioms \ref{A2} and \ref{A3}. Thus if we accept Axiom \ref{A1}, then Axioms \ref{A2} and \ref{A3} follow automatically. We still list them as separate axioms because Theorems \ref{thm2} and \ref{thm3} depend only on Axiom \ref{A2} and Axiom \ref{A3}, respectively. In the next section, we deduce Theorems \ref{thm1}, \ref{thm2}, and \ref{thm3} from these three axioms.
\end{rem}

\section{Main Theorems}\label{main thm}

\begin{lem}\label{lem}
Let $D$ be an oriented link diagram and let $c(D)$, $s(D)$ and $n(D)$ be the number of crossings of $D$, the number of Seifert circles and
 the number of link components, respectively. Then we have

$$s(D) \leq c(D) + n(D)$$

\begin{proof}
For a link diagram $D$, let $S$ be the corresponding Seifert surface obtained from the Seifert algorithm. By shrinking the disks bounded by Seifert circles into vertices and the half twisted bands into edges, we can easily compute the Euler characteristics of $S$, namely $\chi(S) = s(D) - c(D).$ Then we attach a disk to $S$ along each component of the boundary of $S$ to obtain a closed surface which we denote by $T$. Clearly the number of disks that we need to attach is $n(D).$ Note that $T$ may not be connected and the number of components is equal to the number of components of D as a planar diagram, which is less than or equal to $n(D).$ Since each component of $T$ has Euler characteristics at most $2$, therefore $\chi(T) \leq 2n(D)$. On the other hand, $\chi(T) = \chi(S) + n(D) = s(D) - c(D) + n(D).$ Then the inequality in the lemma follows.
\end{proof}
\end{lem}

\begin{rem}
For knots, a better bound holds: let $g(D)$ be
genus of the Seifert surface from the Seifert algorithm.  Then
$ c(D) - s(D) - n(D) + 2 - 2g(D) = 0.$
In particular, $s(D) \leq c(D) + 2.$
\end{rem}

Axiom \ref{A1} implies Theorem \ref{thm1}.

\begin{thm} \label{thm1}
If $r \geq 5$ is an integer not equal to $6$, then given any two positive numbers $\A$ and $\B$, there exists a link diagram $D$ such that, if $D' \sim_r D$, and $D'$ is the trace closure of some braid, then
$$
g(D') > g(D) + \A(\log g(D) + \log c(D))  \qquad \textrm{unless}
$$

$$
dist_r(D,D') > (g(D)c(D))^\B.
$$

\begin{rem}

This is reminiscent of Theorem $A$ in \cite{Freedman}, which basically says a diagram cannot be made  logarithmically  thin via polynomially many operations. Theorem \ref{thm1} refines Theorem $A$ in \cite{Freedman} in the sense that if the resulting diagram has the nice form of a trace closure, we get a much better linear lower bound.
\end{rem}
\begin{proof}

We will show that the failure of this theorem contradicts Axiom \ref{A1}. Assuming the theorem does not hold, then we have:

$\exists \A,\B > 0$ , $\forall $ link diagram $D$, $\exists D' = \hat{\sigma'}^{tr}$ for some braid $\sigma' , \  D' \sim_r D$, such that
$$
dist_r(D,D') \leq (g(D)c(D))^\B \qquad \textrm{and} \qquad g(D') \leq g(D) + \A(\log g(D) + \log c(D))
$$

Translated into complexity theory language, this is essentially to say there exists an algorithm which solves a problem in $NP\subset NDQC1$. This algorithm, with a diagram $D$ as input, outputs a diagram $D'=\hat{\sigma'}^{tr}$ such that $D' \sim_r D$ and $g(D') \leq g(D) + \A(\log g(D) + \log c(D)).$

Given a braid $\sigma \in B_{2n}, |\sigma| = m$, then we know that $g(\hat{\sigma}^{plat}) = 2n, $ and $c(\hat{\sigma}^{plat}) = m.$ By the statement above, there exists a diagram $D' = \hat{\sigma'}^{tr}, \ D' \sim_r \hat{\sigma}^{plat},$ such that $dist_r(\hat{\sigma}^{plat},D') \leq (2nm)^\B,$ and $g(D') \leq 2n + \A(\log 2n + \log m).$

Assume $\sigma' \in B_{n'},$ then $g(D') = 2n'$. So we have the inequality:
$$
n' \leq n + \frac{\A}{2}(\log 2n + \log m).
$$

Notice that $|\sigma'|$ is $poly(n,m)$ since $dist_r(\hat{\sigma}^{plat},D') \leq (2nm)^\B.$

Now we apply the algorithm in Theorem \ref{DQC1} to $\sigma'$. Set the error threshold $ \varepsilon' = \frac{\varepsilon}{d^{n'-n}},$ and note that $d^{n'-n} \leq (2nm)^{\frac{\A}{2}}$. Then we get a circuit $U$ of size $poly(n', |\sigma'|, \frac{d^{n'-n}}{\varepsilon}) = poly(n,m, \frac{1}{\varepsilon}),$ and

$$
Pr\bigg\{\big|\frac{|J(\hat{\sigma'}^{tr};r)|}{d^{n'}} - Z(\sigma')\big | < \frac{\varepsilon}{d^{n'-n}}\bigg\} > \frac{3}{4}.
$$

The above inequality is equivalent to

$$
Pr\bigg\{\big |\frac{|J(\hat{\sigma}^{plat};r)|}{d^{n}} - d^{n'-n}Z(\sigma')\big | < \varepsilon\bigg\} > \frac{3}{4}.
$$

The number $d^{n'-n}$ is efficiently computable on a one clean qubit machine.
Therefore, approximating $\frac{|J(\hat{\sigma}^{plat};r)|}{d^n}$ is a problem in $NDQC1.$ By Theorem \ref{BQP}, this problem is complete in $BQP.$ So it follows that $BQP \subset NDQC1,$ which contradicts Axiom \ref{A1}.

\end{proof}
\end{thm}

\begin{cor} \label{cor1}
If $r \geq 5$ is an integer not equal to $6$, then given any two positive numbers $\A$ and $\B$, there exists a link diagram $D$ such that, for any diagram $D',$  $D' \sim_r D$, we have
$$
c(D') > \frac{g(D)}{2} + \A(\log g(D) + \log c(D)) - n(D) \qquad \textrm{unless}
$$

$$
dist_r(D,D') > (g(D)c(D))^\B,
$$
where $n(D)$ is the number of link components of $D$.

\begin{proof}

Assuming that the corollary is false, we have:

$\exists \A,\B > 0$, $\forall $ link diagram $D$, $\exists D',$ $D' \sim_r D$, such that
$$
dist_r(D,D') \leq (g(D)c(D))^\B \qquad \textrm{and} \qquad
$$
$$
c(D') \leq \frac{g(D)}{2} + \A(\log g(D) + \log c(D)) - n(D).
$$

By Lemma \ref{lem} , $\sharp$(Seifert circles of $D'$) $\leq c(D') + n(D').$

Clearly, the $\Gamma_{r}$-Dehn surgeries preserve the number of components of a link. Thus $n(D')$ is equal to $n(D).$

It's well-known that there exists a classically efficient algorithm to transform a link diagram into the trace closure of some braid diagram while preserving the link type (e.g. see \cite{Yamada}). Moreover, as the algorithm described in Theorem $\textrm{1-1}$ of \cite{Vogel}, if a link diagram has $n$ Seifert circles, then at most $n^2$ Reidemeister II moves suffice to implement this transformation. No other types of Reidemeister moves are needed, and this algorithm preserves the number of Seifert circles.

Applying this algorithm to the diagram $D'$ results the trace closure of some braid $\sigma' \in B_{n'}.$ Since $\sharp$(Seifert circles of $D'$) = $\sharp$(Seifert circles of $\hat{\sigma'}^{tr}$) = $n'$ = $\frac{g(\hat{\sigma'}^{tr})}{2}$, we have
$$
g(\hat{\sigma'}^{tr}) \leq 2c(D') + 2n(D') \leq g(D) + 2\A(\log g(D) + \log c(D))
$$

and
$$
dist_r(D,\hat{\sigma'}^{tr} ) \leq (g(D)c(D))^\B + (\frac{g(D)}{2}+ \A(\log g(D) + \log c(D)))^2
$$

Clearly, $\hat{\sigma'}^{tr} \sim_r D.$ This contradicts Theorem \ref{thm1}.

\end{proof}
\end{cor}

The following theorem appeared in \cite{Freedman}, where it followed from the assumption that $\sharp P \nsubseteq NP.$ Here we deduce it from Axiom \ref{A2}.

Axiom \ref{A2} implies Theorem \ref{thm2}.

\begin{thm}\label{thm2}
If $r \geq 5$ is an integer not equal to $6$, then given any two positive numbers $\A$ and $\B$, there exists a link diagram $D$ such that if $D' \sim_r D$, then
$$
g(D') > \A(\log g(D) + \log c(D)) \qquad \textrm{unless}
$$

$$
dist_r(D,D') > (g(D)c(D))^\B.
$$

\begin{proof}
As in the original proof in \cite{Freedman}, if the theorem is not true, then evaluating the Jones polynomial of a link diagram at the $r$-th root of unity is a problem in $NP$. Then approximating $\frac{|J(\hat{\sigma}^{plat};r)|}{d^n}$ for a braid $\sigma \in B_{2n}$ is a problem in $NP$, which implies $BQP \subset NP,$ contradicting Axiom \ref{A2}.
\end{proof}
\end{thm}

The following corollary is clearly weaker than Corollary \ref{cor1}. Since it follows directly from Axiom \ref{A2} (and Theorem \ref{thm2}), which is weaker than Axiom \ref{A1}, we still include it here with a proof:

\begin{cor}\label{cor2}
If $r \geq 5$ is an integer not equal to $6$, then given any two positive numbers $\A$ and $\B$, there exists a link diagram $D$ such that, if $D' \sim_r D$, then
$$
c(D') > \A(\log g(D) + \log c(D)) \qquad \textrm{unless}
$$

$$
dist_r(D,D') > (g(D)c(D))^\B.
$$

\begin{proof}

If the statement is not true, then $\exists \A,\B > 0$, given any braid $\sigma \in B_{2n}, |\sigma| = m,$ there exists a link diagram $D$, such that $\hat{\sigma}^{plat} \sim_r D, \ dist_r(\hat{\sigma}^{plat},D) < (2nm)^\B $ and $c(D) < \A(\log 2n + \log m)$.

Classically evaluating the Jones polynomial of the diagram $D$ has the complexity $O(2^{c(D)}) < O(2nm)^\A)$. Thus evaluating $\frac{|J(\hat{\sigma}^{plat};r)|}{d^n}$ for a braid $\sigma \in B_{2n}$ is a problem is $NP$, which contradicts to Axiom \ref{A2}.

\end{proof}

\end{cor}

Theorem \ref{thm3} below and its corollary \ref{cor3} are weaker than Theorem \ref{thm1} and Corollary \ref{cor1}.  However, we still point them out separately since they follow from weaker axioms.

Axiom \ref{A3} implies Theorem \ref{thm3}.
\begin{thm} \label{thm3}
If $r \geq 5$ is an integer not equal to $6$ and $\A > 0$, let $Q(r,\A)$ be such a problem which, given a link diagram $D$, computes a braid diagram $\sigma$, such that $D$ and $\hat{\sigma}^{tr}$ have the same Jones polynomial at the $r$-th root of unity and
$$
g(\hat{\sigma}^{tr}) \leq g(D) + \A(\log g(D) + \log c(D)).
$$

Then there is no efficient $DQC1$ algorithm to solve $Q(r,\A)$, i.e. $Q(r,\A)$ cannot be solved in polynomial time by a $DQC1$ machine.

\begin{proof}
Assume there is such an algorithm, then apply it to the plat closure of the braid $\sigma \in B_{2n}, \ |\sigma| = m$. Let $\sigma' \in B_{n'}$ be the output. Then
$$
g(\hat{\sigma'}^{tr}) \leq 2n + \A(\log 2n + \log m).
$$

As in the last part of the proof in Theorem \ref{thm1}, we can approximate $\frac{|J(\hat{\sigma}^{plat};r)|}{d^n}$ by applying the algorithm in Theorem \ref{DQC1} to $\sigma'$ to approximate $\frac{|J(\hat{\sigma'}^{tr};r)|}{d^{n'}}$. By Theorem \ref{BQP}, approximating $\frac{|J(\hat{\sigma}^{plat};r)|}{d^n}$ is $BQP$ complete. This implies $BQP \subset DQC1$, which contradicts Axiom \ref{A3}.

\end{proof}
\end{thm}

\begin{cor}\label{cor3}
If $r \geq 5$ is an integer not equal to $6$ and $\A > 0$, let $R(r,\A)$ be such a problem which, given a link diagram $D$, computes a link diagram $D'$, such that $D' \sim_r D$ and
$$
c(D') \leq \frac{g(D)}{2} + \A(\log g(D) + \log c(D)) - n(D).
$$

Then there is no efficient $DQC1$ algorithm to solve $R(r,\A)$.
\begin{proof}
The proof is basically the same as that of Corollary \ref{cor1}.
\end{proof}
\end{cor}

\bibliographystyle{plain}
\bibliography{mybib}

\begin{thebibliography}{10}

\bibitem{aharonov2011}
Dorit Aharonov and Itai Arad.
\newblock The {BQP-hardness} of approximating the {Jones} polynomial.
\newblock {\em New Journal of Physics}, 13(3):035019, 2011.
\newblock (arXiv:quant-ph/0605181).

\bibitem{Aha09}
Dorit Aharonov, Vaughan Jones, and Zeph Landau.
\newblock A polynomial quantum algorithm for approximating the {Jones}
  polynomial.
\newblock {\em Algorithmica}, 55(3):395--421, 2009.
\newblock (arXiv:quant-ph/0511096).

\bibitem{BFLW}
M.~Bordewich, M.~Freedman, L.~Lovasz, and D.~Welsh.
\newblock Approximate counting and quantum computation.
\newblock {\em Combinatorics Probability and Computing}, 14(5/6):737--754,
  2005.
\newblock (arXiv:0908.2122).

\bibitem{Shepherd}
Shepherd Dan.
\newblock Computation with unitaries and one pure qubit.
\newblock {\em arXiv:quant-ph/0608132}, 2006.

\bibitem{Freedman}
M.~H. Freedman.
\newblock Complexity classes as mathematical axioms.
\newblock {\em Annals of Mathematics}, 170(995), 2009.
\newblock (arXiv:0810.0033).

\bibitem{Freedman03}
Michael Freedman, Alexei Kitaev, Michael Larsen, and Zhenghan Wang.
\newblock Topological quantum computation.
\newblock {\em Bulletin of the American Mathematical Society}, 40(1):31--38,
  2003.
\newblock (arXiv:quant-ph/0101025).

\bibitem{freedman20022}
Michael~H Freedman, Alexei Kitaev, and Zhenghan Wang.
\newblock Simulation of topological field theories by quantum computers.
\newblock {\em Communications in Mathematical Physics}, 227(3):587--603, 2002.
\newblock (arXiv:quant-ph/0001071).

\bibitem{freedman2002}
Michael~H Freedman, Michael Larsen, and Zhenghan Wang.
\newblock A modular functor which is universal for quantum computation.
\newblock {\em Communications in Mathematical Physics}, 227(3):605--622, 2002.
\newblock (arXiv:quant-ph/0001108).

\bibitem{Gilmer}
P.~M. Gilmer.
\newblock Congruence and quantum invariants of 3-manifolds.
\newblock {\em Algebr. Geom. Topol}, 7:1767--1790, 2007.
\newblock (arXiv:math/0612282).

\bibitem{JW}
S.~P. Jordan and P.~Wocjan.
\newblock Estimating {Jones} and {HOMFLY} polynomials with one clean qubit.
\newblock {\em Quantum Information and Computation}, 9(264), 2009.
\newblock (arXiv:0807.4688).

\bibitem{KL}
E.~Knill and R.~Laflamme.
\newblock Power of one bit of quantum information.
\newblock {\em Physical Review Letters}, 81(25), 1998.
\newblock (arXiv:quant-ph/9802037).

\bibitem{Kuperberg09}
Greg Kuperberg.
\newblock How hard is it to approximate the {Jones} polynomial?
\newblock {\em arXiv:0908.0512}, 2009.

\bibitem{Lackenby}
M.~Lackenby.
\newblock Fox's congruence classes and the {quantum-SU (2)} invariants of links
  in 3-manifolds.
\newblock {\em Commentarii Mathematici Helvetici}, 71(1):664--677, 1996.

\bibitem{NS}
S.~H. Ng and P.~Schauenburg.
\newblock Congruence subgroups and generalized {Frobenius-Schur} indicators.
\newblock {\em Communications in Mathematical Physics}, 300(1), 2010.
\newblock (arXiv:0806.2493).

\bibitem{SCHLEIMER}
S.~Schleimer.
\newblock Sphere recognition lies in {NP}.
\newblock {\em arXiv:math/0407047}, 2004.

\bibitem{SJ}
P.~W. Shor and S.~P. Jordan.
\newblock Estimating {Jones} polynomials is a complete problem for one clean
  qubit.
\newblock {\em Quantum Information and Computation}, 8(681), 2008.
\newblock (arXiv:0707.2831).

\bibitem{TR}
V.~G. Turaev and N.~Reshetikhin.
\newblock Invariants of 3-manifolds via link polynomials and quantum groups.
\newblock {\em Invent. Math}, 103(3), 1991.

\bibitem{Vertigan}
D.~Vertigan.
\newblock The computational complexity of {Tutte, Jones, HOMFLY and Kaufman}
  invariants.
\newblock {\em DPhil Thesis, Oxford University, Oxford, England}, 1991.

\bibitem{Vertigan2}
D.~Vertigan.
\newblock The computational complexity of {Tutte} invariants for planar graphs.
\newblock {\em SIAM J. Comput.}, 35(3), 2005.

\bibitem{Vogel}
P.~Vogel.
\newblock Representation of links by braids: A new algorithm.
\newblock {\em Comment. Math. Helvetici}, 65(104), 1990.

\bibitem{Wang}
Zhenghan Wang.
\newblock {\em Topological Quantum Computation}, volume 112.
\newblock Amer Mathematical Society, 2010.

\bibitem{WY}
P.~Wocjan and J.~Yard.
\newblock The {Jones} polynomial: Quantum algorithms and applications in
  quantum complexity theory.
\newblock {\em Quantum Information and Computation}, 0(0), 2003.
\newblock (arXiv:quant-ph/0603069).

\bibitem{Yamada}
S.~Yamada.
\newblock The minimal number of {Seifert} circles equals the braid index of a
  link.
\newblock {\em Invent. Math}, 89(347), 1987.

\end{thebibliography}

\end{document}